\documentclass{aa}
\usepackage{txfonts}
\usepackage[T1]{fontenc}
\usepackage[utf8]{inputenc}
\usepackage{lmodern}
\usepackage{url}
\usepackage{graphicx}
\graphicspath{ {files/} }
\usepackage{float}
\usepackage{comment}
\usepackage{wasysym}
\usepackage{upgreek}

\usepackage{xargs}                      
\begin{document}

\title{Recovering saturated images for\\high dynamic Kernel-Phase analysis}
\subtitle{Application to the determination of dynamical masses for the system Gl 494AB}
\titlerunning{Recovering saturated images for kernel-phase analysis}
\author{R. Laugier\inst{1}\and
	F. Martinache\inst{1}\and
	A. Ceau\inst{1}\and
    D. Mary\inst{1}\and
    M. N'Diaye\inst{1}\and
    J.-L. Beuzit\inst{2}}
\institute{Université Côte d'Azur, Observatoire de la Côte d'Azur, CNRS, Laboratoire Lagrange, France\and
	Aix Marseille Univ, CNRS, CNES, LAM, Marseille, France}


\abstract
{Kernel-phase observables extracted from mid- to high-Strehl images are proving to be a powerful tool to probe within a few angular resolution elements of point sources. The attainable contrast is however limited by the dynamic range of the imaging sensors. The Fourier interpretation of images with pixels exposed beyond the saturation has so far been avoided. We show that in cases where the image is dominated by the light of a point source, we can use an interpolation to reconstruct the otherwise lost pixels with an accuracy sufficient to enable the extraction of kernel-phases from the patched image. We demonstrate the usability of our method by applying it to archive images of the Gl 494AB system, taken with the Hubble Space Telescope in 1997. Using this new data point along with other resolved observations and radial velocity measurements, we produce improved constraints on the orbital parameters of the system, and consequently the masses of its components.}

\maketitle

\section{Introduction}
  The quest for the direct imaging of exoplanets has accelerated in the recent years with the commissioning of extreme adaptive optics systems with coronagraphic capabilities, such as SPHERE \citep{Beuzit2008}, GPI \citep{Macintosh2014} and SCExAO \citep{Jovanovic2015a}. These instruments were designed to achieve high contrast detections ($ 10^{4} $) at small angular separations (down to $ \sim 2 \lambda/D $ where $ \lambda/D $ defines the resolution element, $\lambda$ is the wavelength, and $D$ is the  telescope diameter). The planets discovered thus far by these instruments have however been detected at larger separations: \citet{Macintosh2015} report a detection at $ \sim 10 \lambda/D$ (449 mas) and \citet{Chauvin2017} at $ \sim 20 \lambda/D $ (830 mas). The performance of these coronagraphic devices at small angular separations is currently limited by the quality of the wavefront correction of the adaptive optics system, to which they are extremely sensitive \citep{Guyon2006}.\par
  In the small angular separation regime, the wavefront quality requirements become so stringent that light leakage induced by instrumental phase dominates over the shot noise in the coronagraphic images. In practice, the use of interferometric techniques designed to be robust to the wavefront errors, like closure phases \citep{Jennison}, or kernel-phases (KP) \citep{Martinache2010} are a competitive tool for discovering faint companions around nearby stars. The KERNEL project aims at developing observation and image processing means that provide observables which are intrinsically robust to small wavefront errors and can therefore be used to recover information in the image.\par
  Kernel-phase observables \citep{Martinache2010} provide a reliable probe for the detection of asymmetric features at the smallest separations (down to $ 0.5\lambda/D $) with moderate contrast (80-200:1). The attainable contrast is constrained by the dynamic range of the camera used. The acquisition of images for their kernel-phase analysis requires a compromise: the image must be sufficiently exposed to obtain satisfactory signal to noise ratio from the faint feature, while avoiding saturation on its brightest parts.\par
  By destroying linearity of the intensity distribution, saturation (or clipping) breaks some of the prerequisites of the Fourier transform. As a result the shift theorem and object-image convolution relationship become unusable, preventing the interpretation of the phase and the construction of kernels. Although the limits were pushed both by \cite{Pope2016} who had to linearize the sensor's response (soft saturation) and \cite{Martinache2016a} where the uv plane sampling was truncated to avoid the problematic signals and function in a degraded mode, no kernel-phase analysis has yet been published based on hard saturated images.\par
  The work presented in this paper attempts to circumvent this limitation and extract kernel-phases from images featuring a saturated core after using a saturation recovery algorithm. As a demonstration, we apply this method to the reduction of images taken with the Near Infrared Camera and Multi-Object Spectrograph (NICMOS) on the Hubble Space Telescope (HST) of Gl 494 (DT Vir, LHS 2665, HIP 63510, BD +13 2618) acquired in 1997 which leads to a new detection and measurement of a known companion. The novel visual astrometry data point, combined with available resolved observations at other epochs together with radial velocity observations, lead to improved constraints on the orbital parameters of the system and therefore on the mass of its components.

\section{Interpolation of saturated images and kernel analysis}
  Typical algorithms for bad pixels correction assume that problematic pixels are few and isolated, allowing recovery by interpolating the neighboring pixels in the image, or by minimizing the power associated to the highest spatial frequencies in the Fourier plane \citep{Ireland2013a}. In the case of saturation, the pixels to recover are clustered. Our approach tackles this problem in the image plane, and uses a synthetic Point Spread Function (PSF) as a reference to interpolate the value of the saturated pixels, after which Fourier-phase can then be reliably extracted from these enhanced images.
\subsection{Saturation recovery algorithm}\label{algo}
The restoration of saturated pixels assumes\label{hypo} that the local signal is dominated by the flux of a bright point source. This makes it possible to fit a parametric theoretical (or empirical) PSF to the non-saturated parts of the image and to replace the saturated pixels by interpolation of this adjusted PSF. The best solution of position $(x,y)$ and amplitude $z$ minimizes the following $\chi^2$ variable:
  
  \begin{equation}\label{equ_chi2}
  	\chi^{2}(x,y,z) = || \mathbf{i} - z \cdot \mathbf{p}(x,y) ||^{2} ,
  \end{equation}
\noindent
where $\mathbf{i} = \boldsymbol{\Sigma}^{-\frac{1}{2}} \cdot \mathbf{s}_{img}$ and $\mathbf{p}(x,y) = \boldsymbol{\Sigma}^{-\frac{1}{2}} \cdot \mathbf{s}_{PSF}(x,y)$ are, respectively, error-normalized (whitened) vectors of the image signal $\mathbf{s}_{img}$, and the $(x,y)$ shifted PSF signal $ \mathbf{s}_{PSF}(x,y) $. The synthetic PSF is obtained with the TinyTim \citep{Krist2011} simulator, and both vectors exclude saturated pixels, which ensures the fitting is not biased by their arbitrary value. A centroid algorithm provides a starting point within a few pixels of the final result, scale at which the problem is convex. Minimizing $ \chi^{2}(x,y,z) $ should give the best performance assuming a good estimator of the covariance matrix $\boldsymbol{\Sigma}$ is used.\par
  This covariance has contributions from the image sensor, and from the PSF model. For the image, the pixels are independent (and the contribution diagonal), and a good estimator can be built from the data in the reduced image file, with Eq. \ref{eq:ERR} for the terms on the diagonal. For the PSF however, the deviations in the image plane are strongly correlated, and their estimations would rely on strong hypotheses on the spectral distribution of the aberration modes of the wavefront errors. In the case of a space telescope, thanks to the good stability of the PSF, one would expect to be able to neglect this error term; effectively using the sensor noise as the only source of deviation between the image and the PSF. However it was found that a uniform error estimator made the interpolation less sensitive to the evolution of the small instrumental phase over long timescales, and therefore produced smaller biases in the observables. In practice, since the images are large, an exponential windowing function (sometimes called super-Gaussian) was used to exclude the pixels that have low SNR:
  \begin{equation}
     g = e^{-(\frac{r}{r_{0}})^4},
  \end{equation}
  where $r$ is the distance from the pixel to the approximated center of the star, and $r_{0}$ is a radius parameter chosen depending on the SNR of the image. From the error estimation point of view this can be seen as using the inverse of the values of the mask as the corresponding $\sigma^2_{ii}$ terms of a diagonal matrix $\mathbf{\Sigma}$. A constant value of $r_{0} = 40$ pixels gave satisfactory results with the NICMOS images and was used for the whole dataset.\par
An analytical expression of $\hat{z}$, the optimal value of $z$ can be obtained from Eq. \ref{equ_chi2}:
  \begin{equation}\label{zhat}
  	\hat{z}(x,y) = \frac
      {\mathbf{p}(x,y)^{t} \cdot \mathbf{i}}
      {\mathbf{p}(x,y)^{t} \cdot \mathbf{p}(x,y)} ,
  \end{equation}
leading, through the substitution in Eq. \ref{equ_chi2} to the definition of a new criterion, function of x and y only:
  \begin{equation}
  	\varepsilon(x,y) = - \frac
      {(\mathbf{p}(x,y)^{t} \cdot \mathbf{i})^2}
      {\mathbf{p}(x,y)^{t} \cdot \mathbf{p}(x,y)}.
  \end{equation}
  \par
Minimizing $\varepsilon(x,y)$ is therefore equivalent to minimizing $\chi^2(x,y,z)$ but is computationally more efficient.

\subsection{Simulation of realistic} NICMOS images\label{simulation}
  The simulation of images is necessary for two purposes: the evaluation of the fidelity of the algorithm in Sect. \ref{interpretation}, and the bootstrapping of the kernel-phase covariance matrix in Sect. \ref{covariance}.\par
%
%
  Images of single and binary stars were simulated based on a PSF of HST obtained with the TinyTim software. Binary stars were constructed by adding a shifted and contrasted copy of the original PSF. The noise behavior of the non-destructive reads was also emulated in order to reproduce the typical behavior of the sensor in MULTIACCUM/STEP128 mode. First the number of correct readouts and exposure time is estimated for each pixel through an iterative process using the full well capacity and the flux. Then the readout noise map is estimated as:
  \begin{equation}
  	\sigma_{ro} = \frac{\sigma_{ro,single}}{\sqrt{N_s}},
  \end{equation}
\noindent
where $\sigma_{ro,single}$ is the readout noise value for a single read and $N_s$ is the number of successful samples. This is useful when simulating images in order to build sensible approximations for the metadata maps included in the cal.fits files, but isn't necessary when bootstrapping the errors for existing data as in Sect. \ref{covariance}.\par
\begin{figure}
    \begin{center}
      \includegraphics[width=0.45\textwidth]{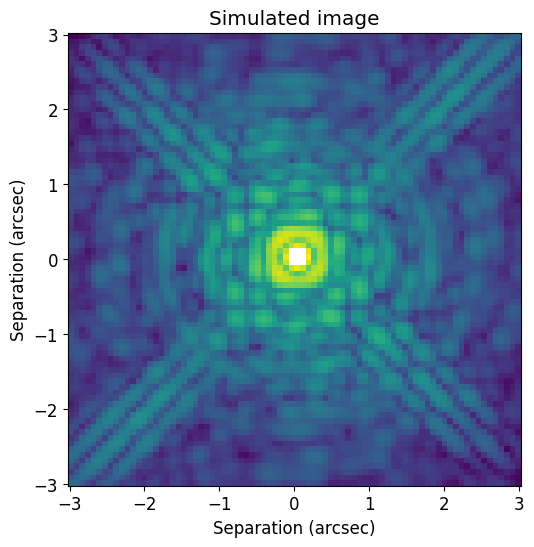}
      \caption{Example of a simulated HST NICMOS2 128 seconds image through the F222M filter ($2.22\mu m$) including shot, readout noise and saturation. The plate scale is 76.5 mas per pixel and the image also features a companion at $10^{3}$ contrast located 450mas left and 450mas down from the primary. The clipped value of nine pixels at the core of the PSF (displayed in white) would usually prevent the use of Fourier analysis.}
      \label{fig:image_example}
    \end{center}
  \end{figure}

  For the application of photon (shot) noise, we follow the directions (and notations) provided in the NICMOS data handbook \citep{Thatte} and consider a total error map of:
   \begin{equation}\label{eq:ERR}
   	 ERR_{total} = \sqrt[]{ \frac{SCI}{G\times TIME} + ERR^{2} },
   \end{equation}
\noindent
where $SCI$ is the flux in count per second, G is the ADC \textit{inverse} gain (in electron/ADU), $ERR$ is the noise map, and $TIME$ is the map of the total exposure time, all available in the FITS file. Figure \ref{fig:image_example} shows an example of a simulated binary star image.\par
  The images are then clipped to a maximum value, interpolated using the algorithm described in Sect. \ref{algo}, and compared to ideal (non saturated) images to judge of the fidelity of the algorithm.\par

\subsection{The Kernel-Phase method and pipeline}\label{KP}
\begin{figure}
  \centering
      \includegraphics[width=0.45\textwidth]{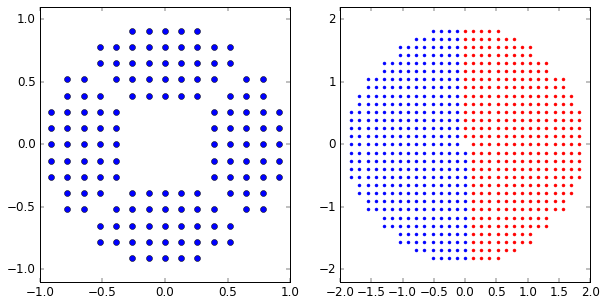}
      \caption{Discrete representation of the HST for kernel analysis of the NICMOS images. The left panel is a representation of the pupil model defined by the cold mask of the instrument. The model takes into account central obstruction and spider vanes, but neglects finer structures such as support pads. The grid pitch is 0.13 m, outer diameter is 1.95 m, inner diameter is 0.71 m and thickness of the spider vanes is 0.07 m. The right panel shows the corresponding uv-plane coverage.}
      \label{fig:pupil_podel}
  \end{figure}

  Kernel analysis relies on the discrete model of the pupil of the telescope which was built using a dedicated python package called \verb+XARA+
  \footnote{XARA is available at \url{github.com/fmartinache/xara}}, developed in the context of the KERNEL project. 
A discrete pupil model of 140 subapertures was constructed following a Cartesian grid covering the aperture of the telescope with an outer diameter of 1.95 m, with an inner obstruction diameter of 0.71 m and spider 0.07 m thickness spider arms. Figure \ref{fig:pupil_podel} shows the sampling of the model both in the pupil plane and in the UV plane. According to the Nyquist-Shannon sampling theorem, the grid's pitch of $b_{min} = 0.13$ m allows for detection up to $\rho_{max} = \lambda/2b_{min} \approx 1.76$ arcsec at $ 2.22 \mu m $ without aliasing.\par

The extraction of kernel-phases requires five steps:
  \begin{enumerate}
    \item Integer pixel recentering of the image.
    \item Application of an apodization mask to reduce the effect of readout noise and avoid Fourier domain ringing.
    \item Sub-pixel recentering of the image by application of a wedge phasor in the Fourier domain.
    \item Extraction of the complex visibility vector.
    \item Multiplication of the phase of the visibility by the Kernel matrix.
  \end{enumerate}

This procedure is applied to each image of a given data-set, and produces each time
a vector $\boldsymbol{\kappa}$ of $n_k=262$ kernel-phases.
  
 \subsection{Fidelity of the saturation recovery procedure} \label{interpretation}
  Since the aim of this study is to enable the extraction of kernel-phases, the main concern is with the fidelity of the phase of the Fourier transform of the image. The discretized pupil model (see Sect. \ref{KP}) provides a comprehensive sampling scheme of the Fourier plane.\par
\begin{figure}
    \begin{center}
      \includegraphics[width=0.24\textwidth]{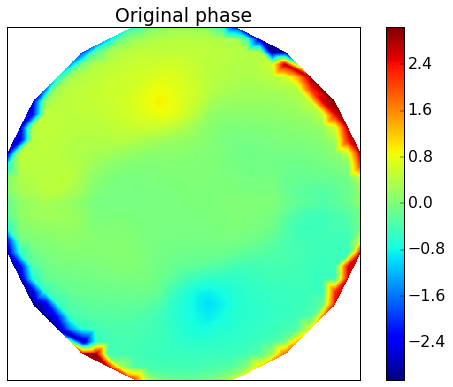} 
      \includegraphics[width=0.24\textwidth]{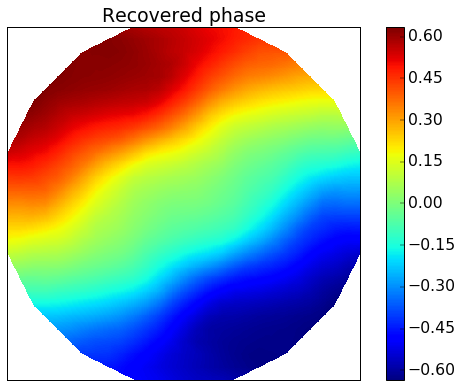}
      
      \includegraphics[width=0.24\textwidth]{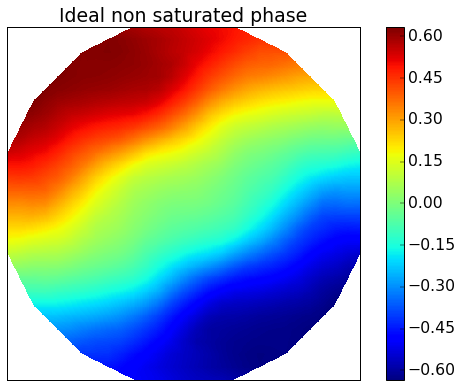} 
      \includegraphics[width=0.24\textwidth]{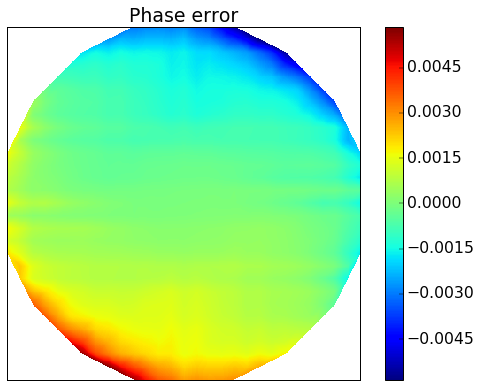}
      \caption{Representation of the Fourier plane phase (in radians) sampled for an example of saturated image of a binary star (shown in Fig. \ref{fig:image_example}), for the ideal not saturated image, for the recovered image and the residual. For the original image (upper left) the saturation results in large amplitude aberrations in the outer ring. After recovery (upper right), the phase is very close to what it would be if the sensor had remained linear (lower left). Difference between the two (lower right) shows small residuals concentrated in the outermost ring.}
      \label{fig:phase_images}
    \end{center}
  \end{figure}

  A binary star image with a $10^{3}$ contrast companion is simulated as described in \ref{simulation} with a central peak intensity corresponding to four times the dynamic range of the sensor, resulting in nine clipped pixels. Figure \ref{fig:phase_images} shows the Fourier phase compared between the saturated image where the outer edge of the uv plane is unusable, the ideal non-saturated image, and the recovered image for which the phase information was restored.
  
\begin{figure}
    \begin{center}
      \includegraphics[width=0.5\textwidth]{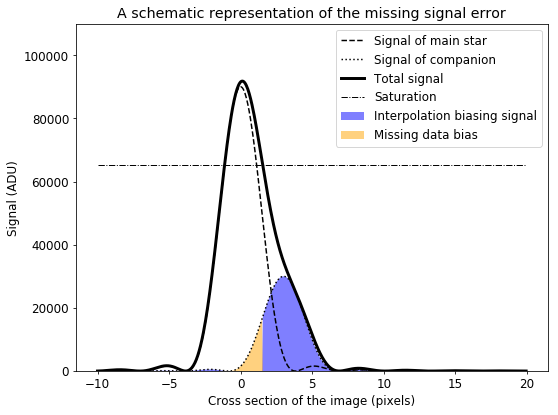}
      \caption{Illustration of both the deviation from the single star PSF that influence the fitting, and the bias of recovered pixel due to omitted feature signal. This configuration at a very low contrast of $3$ and $ \sim 1 \lambda/D $ separation exaggerates the situation.}
      \label{fig:syserror}
    \end{center}
  \end{figure}

  Our recovery algorithm assumes that the image is dominated by the PSF of the primary star. The end result is therefore likely to be affected by the three following potential biases:
\begin{enumerate}
\item \label{sigbias}The signal of the companion acquired by the saturated pixels is lost and omitted by the recovery process as illustrated in orange by Fig. \ref{fig:syserror}.
\item \label{fitbias}The signal of the companion still present in the image, but not in the ideal fitted PSF, therefore biasing the interpolated value; showed in blue in Fig. \ref{fig:syserror}.
\item \label{phasebias}The instrumental PSF is distorted by a small amount of instrumental phase, therefore biasing the interpolated value. The impact of this bias was mitigated through the use of a uniform deviation estimation in section \ref{algo}.
\end{enumerate}

  At high contrast, the effect of bias \ref{sigbias} is negligible compared to shot noise. 
  For bias \ref{fitbias}, it is much more difficult to evaluate as the effect is non-linear and highly dependent on the geometry of the PSF.\par
\begin{figure}
    \begin{center}
      \includegraphics[width=0.5\textwidth]{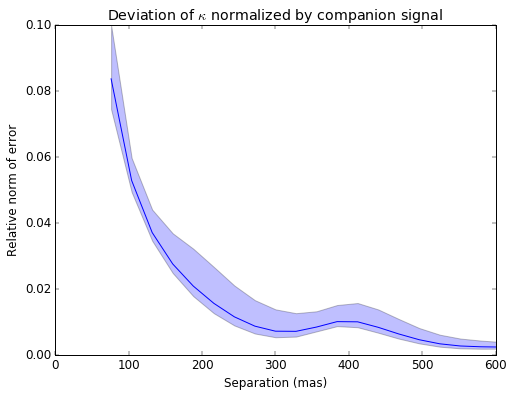}
      \caption{Relative norm of the kernel-phase bias due to saturation-recovery process (normalized by the amplitude of the companion signal) obtained by noiseless simulations. The colored region indicates the range of values for contrasts between 10 and 250, while the solid line indicates their mean. As we can see, the bias quickly drops to below 2\% past the $ \sim 1 \lambda/D = 235$ mas separation.}
      \label{fig:kp_error}
    \end{center}
  \end{figure}

  We characterized the combined effect of biases \ref{sigbias} and \ref{fitbias} on kernel-phases through a controlled noiseless simulation in a case with 4 saturated pixels, representative of the GL 494 case highlighted in Section \ref{gl494}. We measured the amplitude of the relative error $ e = \frac{|| \boldsymbol{\kappa}_e - \boldsymbol{\kappa}_r ||}{|| \boldsymbol{\kappa}_e ||} $, where $ \boldsymbol{\kappa}_e$ is the expected kernel-phase signal, and  $ \boldsymbol{\kappa}_r $ is the recovered kernel-phase signal. In the high contrast regime, this metric sets an upper bound to the fitted contrast, but also generally encompasses fitting residuals in the whole kernel-phase subspace. As shown in Fig. \ref{fig:kp_error}, this metric mostly depends on the separation, and it drops below 2\% at separations $\geq 200$ mas.
\par
  The influence of biases \ref{fitbias} and \ref{phasebias} are both affected by the choice of weighting scheme mentioned in Sect. \ref{algo}. Although the weighting function used produced satisfying results, further optimization work could lead to more improvements in the future.\par

\section{Kernel analysis of NICMOS archival images}
  The dataset that we propose to examine comes from the Hubble Space Telescope proposal \# 7420. The aim of the study was to observe nearby stars within 10pc of the Sun in order to find very low mass ($ M<0.2M $) stars. Observations were conducted using the HST Near Infrared Camera and Multi-Object Spectrograph camera 2 through the F110W, F180M, F207M and F222M filters during from 1997 to 1998.\par
  Analysis of the data was published by \citet{Krist1998} and \citet{Dieterich2012} where PSF subtraction was performed through manual adjustments. They determined detection limits by trying to identify artificially inserted companion stars over the subtraction residual. Contrast detection limits are given for the F180M filter from 0.2'' to 4''. In comparison, our model is sensitive between 0.12'' to 1.76'' due to its extent and resolution. In this work, we focus on the images taken in the F222M filter consisting of 159 images of 80 targets.\par
  
\subsection{Saturated and bad pixels}\label{realidentification}
  The first step of the reduction process is the identification of problematic pixels. When considering the raw output of a camera, this might seem trivial: identify the pixels that have reached the maximum ADC value of the camera, often in the form $ 2^{n_{bits}} $ in scientific sensors. However, this will prove erroneous when the gain of the camera is chosen for maximal dynamic range and the value is actually limited by the full well capacity of the pixel. This value can vary from pixel to pixel and is often more of a soft maximum showing a non-linear region (that is often linearized by advanced reduction pipelines). Furthermore, dark subtraction and flat-fielding will also play a role in making the clipped pixels harder to identify. For all these reasons, it is more appropriate to have the image reduction pipeline identify the saturated pixels and provide the information along with the meta-data. Here we relied on the metadata provided along with the image in the cal.fits output of the CALNICA reduction pipeline. This is explained in more detail for our case in chapter 2.2.1 of the NICMOS Data Handbook by \cite{Thatte}.\par
  All the images of the dataset were obtained using MULTIACCUM mode with a STEP128 sampling scheme for 128 seconds exposures. This is a mode that uses multiple nondestructive readouts to compute the flux on each pixel. If saturation is detected in some of the last samples, the first non-saturated samples can still be used to compute the flux. In this case, a pixel is only considered saturated if the first $0.303$ seconds of exposure are sufficient to saturate the pixel. This explains the remarkable dynamic range of single images.\par
  Bad pixels are also identified during this step. Bad pixels near the core of the PSF would be treated as saturated pixels and recovered, but pixels further out are simply interpolated with their nearest neighbors.

\subsection{Whitening for the image plane errors}\label{whitening}
  Usually in Fourier phase analysis, the image noise (mainly shot and readout noise) translates into correlated phase noise and, in turn, correlated kernel-phase noise. Such correlations must be accounted for before model-fitting or hypothesis testing. De-correlation can be performed using a whitening transformation described by the matrix 
  \begin{equation}
  	\mathbf{W} = \boldsymbol{\Sigma}_{K}^{-\frac{1}{2}},
  \end{equation}
  where $ \boldsymbol{\Sigma}_K $ is an estimate of the covariance of $\boldsymbol{\kappa}$, the kernel-phase signal vector; therefore ensuring that:
  \begin{equation}\label{equ:cov}
  	\text{cov}(\mathbf{W} \cdot \boldsymbol{\kappa}) \approx \mathbf{I}.
  \end{equation}
  This is very similar to the approach of \citet{Ireland2013a} who uses the finite-dimensional spectral theorem to compute a unitary matrix that diagonalizes the covariance, then normalizes the observables by the corresponding standard deviation, which are the square roots of the terms of the diagonal matrix. In our case the same goal is reached through a non-unitary matrix $\mathbf{W}$ that also applies normalization, as shown by Eq. \ref{equ:cov}.\par
  
Using this whitening matrix, the reduced goodness-of-fit parameter $\chi_{\nu}^2$ writes as:
  
  \begin{equation} 
  	\chi_{\nu}^{2} = \frac
      {||\mathbf{W} \cdot \boldsymbol{\kappa}_{o} - 
         \mathbf{W} \cdot \boldsymbol{\kappa}_{c} - 
         \mathbf{W} \cdot \boldsymbol{\kappa}_{m}||^2}
      {n_{k}-n_{p}},
  \end{equation}

\noindent
where $\boldsymbol{\kappa}_o$, $\boldsymbol{\kappa}_c$ and $\boldsymbol{\kappa}_m$ respectively represent the kernel-phase
for the object of interest, the calibrator (explained in \ref{calibration} and the model. $\boldsymbol{\kappa}_m$ is computed
using a parametric binary object model of $n_P = 3$ parameters: angular separation
$\rho$, position angle $\theta$ and contrast $c$, which leaves $\nu = n_k - n_p$ degrees of
freedom at the denominator. The minimization of this $\chi_{\nu}^2$ with a Levenberg-Marquardt algorithm allows us to identify the best fit model.
  
\subsection{Estimation of the covariance of the Kernel-Phase}\label{covariance}
  The evaluation of the covariance matrix $\boldsymbol{\Sigma}_{K}$ of the calibrated kernel-phase vector $ \boldsymbol{\kappa} = \boldsymbol{\kappa}_{o} - \boldsymbol{\kappa}_{c}$ is necessary for the whitening step (\ref{whitening}). It is obtained for each target-calibrator pair as:
  \begin{equation}
    \boldsymbol{\Sigma}_{K} = \text{cov}(\boldsymbol{\kappa}_{o}) +
    							\text{cov}(\boldsymbol{\kappa}_{c}).
  \end{equation}
  \par
  Although those covariances could be approximated analytically for normal images, the saturation recovery process used here is bound to introduce covariance even within the image plane, making the problem more complicated. Building a reliable covariance matrix estimator for the 262 observables vectors requires the acquisition of a large number of realizations. Since the data available only consists of two snapshots, we use a bootstrapping Monte Carlo approach. A number of realizations are simulated by adding noise to the real science image as described in Sect. \ref{simulation}. Those realizations are then pushed through the same pipeline as the science image as described in Sect. \ref{algo} for the recovery and Sect. \ref{KP} for the extraction of the observables. The only exception is the saturation mask which is kept identical. Since this process is rather computationally intensive because of the necessity to interpolate each image, we use 5000 realizations for each target, and the same covariance estimator is used for both images of the same target because they share the same exposure conditions, sensor location, and aberrations.\par
  
\subsection{Colinearity maps}\label{colin_map}
  In addition to the three parameters $\chi_{\nu}^{2}$ used for model fitting, colinearity maps can be built that constitute a matched filter as defined by \citet{Scharf}:
  \begin{equation}\label{eq:colin_maps}
  M(\alpha,\delta) = \frac
  	{ (\mathbf{W} \cdot \boldsymbol{\kappa}_{o} -
    \mathbf{W} \cdot \boldsymbol{\kappa}_{c})^{T}
    \cdot \mathbf{W} \cdot \boldsymbol{\kappa}_{m}(\alpha, \delta, c_{0})}
  	{|| \mathbf{W} \cdot \boldsymbol{\kappa}_{m}(\alpha, \delta, c_{0}) ||},
  \end{equation}
where $c_0$ is a fixed high contrast value ($10^2$ is used in our case). \par
  In the high contrast regime, the uv phase signal $ \boldsymbol{\Phi}_{0} = \mathrm{arg}(1 + \frac{1}{c}e^{-i\frac{2\pi}{\lambda}(u\alpha_0 + v\delta_0)} )$, and therefore of $\mathbf{W} \cdot \boldsymbol{\kappa}$ are inversely proportional to $c$. As a consequence, $  M(\alpha,\delta) $ will peak at coordinates $ [\alpha, \delta] = [\alpha_0, \delta_0]  $, regardless of their contrast. For this reason, this map is useful for 2D visualization of binary signals present in the kernel signature. It can be computed very quickly over a grid, and plotted as in Fig. \ref{fig:colinmaps_calib}.

\subsection{Choice of calibrators}\label{calibration}
The calibration of interferometric observations requires the selection of point sources observed under similar conditions and with spectral and photometric properties similar to those of the target. The dataset associated with proposal \# 7420 consists of images acquired on a wide variety of targets with respect to intrinsic brightness and spectral types; with observations separated by several weeks. We used two different procedures for: the detection of binary signals among all targets of the sample and; for the binary model fitting and parameter estimation.\par
\begin{figure}
    \begin{center}
      \includegraphics[width=0.5\textwidth]{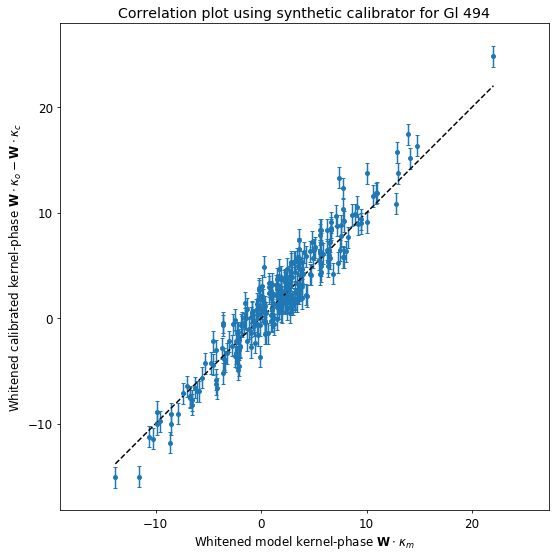}

      \caption{Correlation plot showing the good agreement between the Gl494 data and the fitted binary model. Here the data is whitened and calibrated with the mean of list of 10 selected calibrators. The covariance matrix used for whitening takes into account the covariance of the target, the covariances of the calibrators and the variance between calibrators.}
      \label{fig:correlation}
    \end{center}
 \end{figure}

 
For the detection, we built a generic calibrator through the elimination of images producing outlier kernel signals (in the sense of Euclidean distance to the median of the remaining subset) one at a time until 10 out of the original 159 remained. This conservative approach excludes occurrences of spurious signals caused by potential failures of the interpolation process, sensor defects and the mask alignment inconsistency mentioned by \citet{Krist1998}, as well as targets with resolved features. The mean of the 10 remaining kernel-phase signals constitutes the generic calibration signal, subtracted from all raw extracted kernel-phases, that allowed the identification of the companion in the saturated images of Gl 494, as shown in Fig. \ref{fig:correlation}.\par
\begin{figure}
    \begin{center}
    	\begin{tabular}{ll}
      \includegraphics[width=0.235\textwidth]{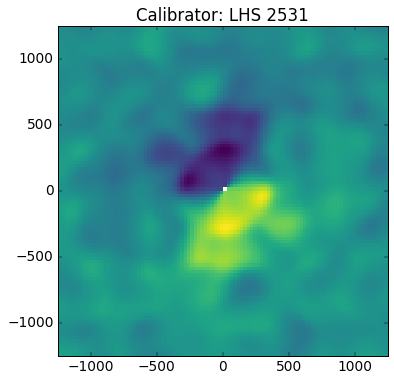} & \includegraphics[width=0.235\textwidth]{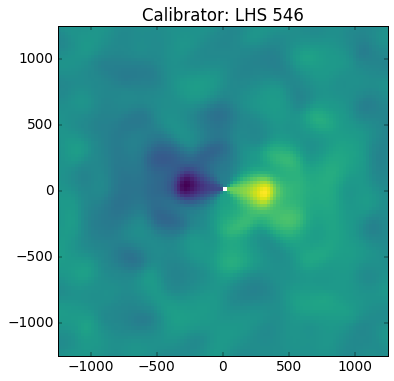}\\
      \includegraphics[width=0.235\textwidth]{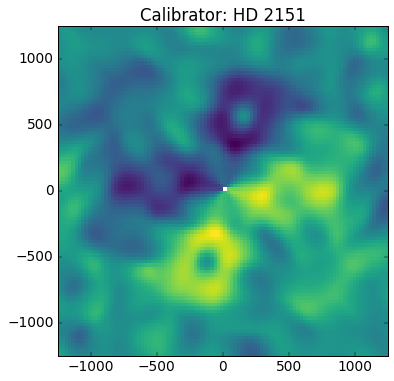} & \includegraphics[width=0.235\textwidth]{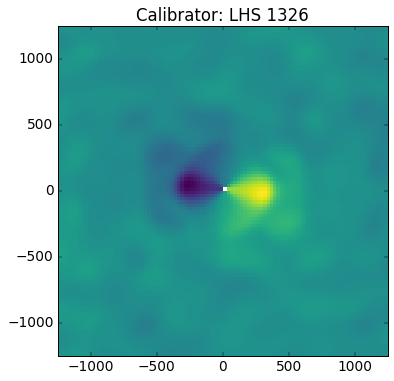}\\
      	\end{tabular}
      \caption{Examples of colinearity maps described in \ref{colin_map} for rejected calibrators (left) where signal from the calibrator is prominent, and selected calibrators (right) where the signal of interest is dominant and visible on the right hand side of the center. In the high contrast approximation, the value of each pixel peaks for calibrated signal vector colinear to the signal of a companion at the pixel's location, regardless of its contrast. Note that the map is antisymmetric, reflecting that kernel-phase is -- like closure phases -- a measure of the asymmetries of the target.}
      \label{fig:colinmaps_calib}
    \end{center}
  \end{figure}

Evaluating the impact of this calibration-induced bias on the extracted binary parameters requires as many calibrators as possible. Suitable ones were selected using visual examination of colinearity maps for Gl 494 after subtraction of each of the 79 other possible calibrators (see Fig. \ref{fig:colinmaps_calib}). Candidates for which the map was not dominated by the binary signal identified at the first stage were discarded, leaving the 32 targets listed in Table \ref{tab:calibrators}. The distribution of the parameters fitted after subtraction of each calibrator signal was used to estimate the confidence intervals for the final measurement.\par

  \begin{table}
\renewcommand{\thetable}{\arabic{table}}
\centering

\caption{List of selected calibrators} \label{tab:calibrators}
\begin{tabular}{ccc}
\hline\hline
Identifier & Spectral type & K\\
        &             &  (mag)\\
\hline 
LHS 1326& M5.5V & 8.93\\
LHS 3558 & M3V & 5.93\\
HD 204961 & M2/3V & 4.50\\
HD 209100 & K5V & 2.24\\
LHS 531 & M3V & 5.81\\
LHS 546 & M5.0Ve & 8.18\\
LHS 4003 & M4.5V & 7.23\\
BD +01 4774 & M1VFe-1 & 5.04\\
LHS 31 & M4V & 6.39\\
HD 42581 & M1V & 4.17\\
HD 260655 & M0.0Ve & 5.86\\
LHS 223 & M5.0Ve & 8.23\\
CD -45 5375 & M1.0 & 5.78\\
HD 85512 & K6Vk: & 4.72\\
LHS 288 & M5.0V & 7.73\\
FI VIR & M4V & 5.65\\
LHS 316 & M7 & 7.64\\
HD 109358 & G0V & 2.72\\
FN VIR & M4.5Ve & 7.66\\
LHS 3233 & M3.0V & 8.05\\
LHS 3255 & M3.5Ve & 7.12\\
BD +25 3173 & M2V & 5.62\\
LHS 3262& M5.0V & 7.92\\
HD 157881 & K7V & 4.14\\
CD -46 11540 & M3V & 4.86\\
BD +18 3421 & M1.5Ve & 5.57\\
HD 165222 & M0V & 5.31\\
LHS 465 & M4.0V & 7.74\\
LHS 476 & M4.0Ve & 7.93\\
HD 190248 & G8IV & 2.04\\
HD 191849 & M0V & 4.28\\
HD 192310 & K2+V & 3.50\\
\hline
\end{tabular}
\end{table}

\section{Gl 494}\label{gl494}
	The binary system Gl 494AB was first identified as an astrometric binary by \citet{Heintz1994} of 14.5 years period. The system was later resolved with the PUEO adaptive optics \citep{Veran1999} at the Canada-France-Hawaii Telescope (CFHT) with a  brief summary of the properties of the system in \cite{Beuzit2004}. The dynamical masses derived remained inconclusive because they relied on the Heintz orbital parameters which are given without standard errors, and show some inconsistency with the position angle. The system was later resolved again with VLT NaCo \citep{Lenzen2003,Rousset2003} in 2005 and 2006 as published by \citet{Ward-Duong2015}, but orbital parameters were only recently updated by \citet{Mann2018a}.\par
\begin{figure}
    \begin{center}
      \includegraphics[width=0.20\textwidth]{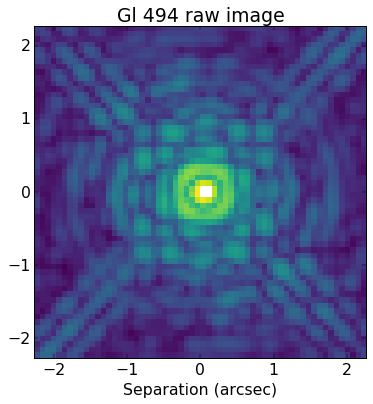}
      \includegraphics[width=0.24\textwidth]{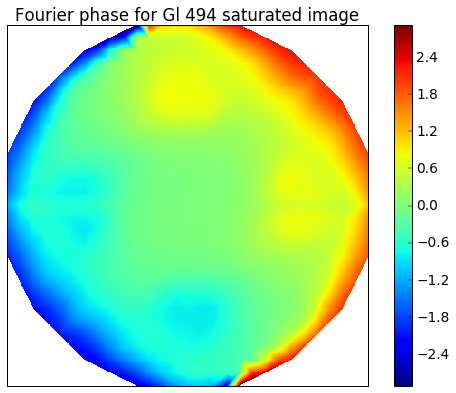}
      \caption{\textit{Left:} Cropped image of Gl 494 taken in F222M filter in log scale before recovery. The four saturated pixels appear in white. The companion is embedded within the first airy ring, to the right of the primary, and not discernible by eye. \textit{Right:} Fourier phase of Gl 494 image with saturation. The spurious phase signal visible in the outer ring would usually prevent interpretation.}
      \label{fig:gl494_im}
    \end{center}
  \end{figure}

  The Hubble Space Telescope archive images of the system used here went overlooked in the early studies, probably for two combined reasons: the system is close to the resolution limit of the telescope ($ \approx 1.3 \lambda/D $) with a high contrast ($ \approx 55 $), and the central peak of the images is saturated. Contrast sensitivities for this target in the same dataset in the F180M filter were later reexamined by \citet{Dieterich2012} to be $\Delta_H$ = 0.1 at 0.2'' and $\Delta_H$ = 2.6 at 0.4'' on this target which is insufficient for this system. Figure \ref{fig:gl494_im} shows a crop of the saturated image used here, and the corresponding unusable uv phase signal. Overcoming the saturation problem allows us to use the more sensitive kernel analysis and adds a valuable high precision visual astrometric point. With the recent observations reported by \citet{Mann2018a}, the visual observations of this binary system now span 18 years, covering more than one orbital period. Together with the radial velocity data, these observations give strong constraints on the orbital parameters of the system, and therefore on the masses of its components.\par

\subsection{Extraction of the visual astrometry}
  The method was applied to images of the system Gl 494 taken with the HST Near Infrared Camera and Multi-Object Spectrograph (NICMOS) camera 2 through the F222M filter in August 1997. This configuration provides 3.1 pixels per resolution element. The kernel-phase vectors of both images were then averaged to reduce noise. The steps described in Sect. \ref{calibration} are followed to obtain a detection, then a distribution of the parameters of the binary which are provided in the first row of table \ref{tab:visual} for use in the orbital determination. The kernel-phase analysis provides a standard deviation 5 mas in separation and $2.3\deg$ in position angle which is significantly lower than what was obtained for the same target with larger ground based telescopes. We also measure the contrast at $\Delta_K = 4.36 \pm 0.15$ which is consistent with the initial PUEO measurement by \citet{Beuzit2004}, and later measurements by \citet{Mann2018a} showing $\Delta_{K_s} = 4.269 \pm 0.017$, but not with the NACO measurements presented by \citet{Ward-Duong2015}. An instrumental bias in the NACO measurements, or the variability of the active primary could explain these differences.\par

\begin{table*}
\renewcommand{\thetable}{\arabic{table}}
\centering
\caption{Visual astrometry data (projected on the plane orthogonal to the line of sight). References: (1) \citet{Ward-Duong2015}; (2) \citet{Mann2018a}. } \label{tab:visual}
\begin{tabular}{ccccc}
\hline \hline 
Epoch          & $\rho$ & $\theta$ &   Contrast & Instrument\\
(decimal year) & (mas) & (deg. EoN) &  (mag)  &             \\
\hline
1997.614 & $295 \pm 5 $    &  $137.8 \pm 2.3$        & $4.36 \pm 0.15$ (F222M) & HST/NICMOS  \\
2000.134 (2) & $475.1\pm7.1$   & $81.4\pm2.8$        & $4.41\pm0.30$ (K)  & CFHT/KIR   \\
2001.337 (2) & $526.3\pm8.2$   & $66.9\pm2.6$        & $4.79\pm0.45$ (BrG)  & CFHT/KIR  \\
2001.515 (2) & $522.0\pm6.2$   & $65.3\pm2.5$        & $4.54\pm0.20$ (BrG)  & CFHT/KIR  \\
2001.592 (2) & $527.6\pm5.0$   & $64.1\pm2.6$        & $4.63\pm0.16$ (BrG)  & CFHT/KIR  \\
2002.170 (2) & $533\pm12$      & $56.2\pm1.2$        & $3.55\pm0.63$ (K)  & Keck/NIRC2 \\
2005.328 (1) & $280 \pm 50$   &   $357.0 \pm 1.0$      & 3.26 ($K_s$)       &  VLT/NaCo \\
2006.389 (1) & $240 \pm 10$  &   $307.4 \pm 4.5$     &  3.71 ($K_s$)      &  VLT/NaCo \\
2006.389 (2) & $236.2\pm3.9$   & $304.62\pm0.96$     & $4.54\pm0.12$  (J) & VLT/NaCo   \\
2006.389 (2) & $233.6\pm5.7$   & $304.3\pm1.5$       & $4.527\pm0.086$ (H) & VLT/NaCo   \\
2007.142 (2) & $270.6\pm8.2$   & $269.3\pm1.1$       & $4.064\pm0.076$ (H) & VLT/NaCo  \\
2009.323 (2) & $309.38\pm0.67$ & $203.182\pm0.069$   & $4.724\pm0.035$ (H) & Keck/NIRC2 \\
2009.323 (2) & $307.5\pm1.1$   & $202.950\pm0.049$   & $4.746\pm0.093$ (J)& Keck/NIRC2 \\
2009.323 (2) & $308.2\pm1.8$   & $203.22\pm0.10$     & $5.46\pm0.17$   (Lp)& Keck/NIRC2 \\
2013.301 (2) & $448.30\pm0.66$ & $86.777\pm0.041$    & $4.359\pm0.013$ (Ks)& Keck/NIRC2 \\
2015.471 (2) & $524.49\pm0.27$ & $60.087\pm0.016$    & $4.301\pm0.012$ (K)& Keck/NIRC2  \\
\hline
\end{tabular}

\end{table*}

\subsection{Determination of orbital parameters}
  In order to determine orbital parameters for the system, we use the radial velocity data obtained with the ELODIE spectrograph \citep{Baranne1996} at Observatoire de Haute-Provence between 2000 and 2006 obtained through the ELODIE archive, and the CORAVEL spectrograph between 1983 and 1995. The standard deviation figure used for ELODIE data was 75 m/s. Two data points acquired with the SOPHIE spectrograph are also available but were not included in the model because it was not worth calibrating for one more instrument.\par
\begin{figure}
    \begin{center}
      \includegraphics[width=0.5\textwidth]{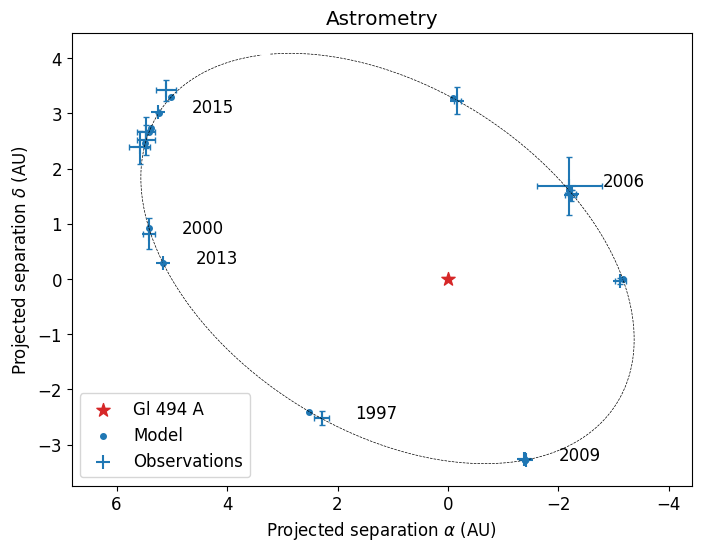}
      \includegraphics[width=0.5\textwidth]{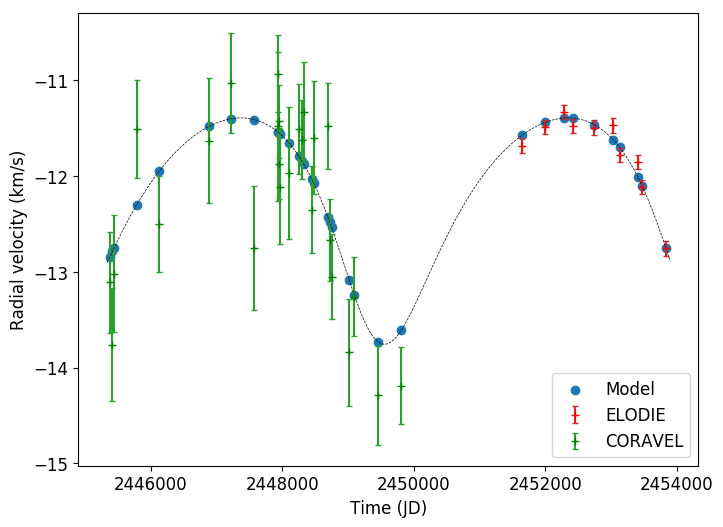}
      \caption{Orbital fitting for the Gl 494AB system. \textit{Upper:} The astrometric measurements and corresponding fitted model results. The 1997 point is the novel measurement provided by this work. \textit{Lower:} The radial velocity measurements obtained with CORAVEL (1983-1995) and ELODIE (2000-2006), and the corresponding fitted model results.}
      \label{fig:orbit}
    \end{center}
  \end{figure}

  The parallax of the system is given by the Gaia DR2 catalog at $ \pi = 86.86 \pm 0.1515 $ mas \citep{GaiaCollaboration2018,GaiaCollaboration2016} which gives a distance of $d = 11.513 \pm 0.02$ pc.
  The technique employed to fit the orbital parameters with both the visual orbit and the radial velocity data is similar to the one described by \citet{Martinache2007}. The \verb+lmfit+ package \citep{Newville}, a python implementation of the Levenberg–Marquardt algorithm is used to obtain a least-square fit for a 10 degrees of freedom problem, resulting in the parameters presented in table \ref{tab:parameters} and Fig. \ref{fig:orbit}.

\begin{table*}
\renewcommand{\thetable}{\arabic{table}}
\centering
\caption{Orbital parameters} \label{tab:parameters}
\begin{tabular}{cccc}
\hline \hline 
Parameter            & \citet{Heintz1994}  & \citet{Mann2018a} & This work \\
\hline
$P$ (years)      &   14.5    & $13.709_{-0.037}^{+0.036}$ &  $13.63 \pm 0.03 $     \\
$T_p$ (decimal year)&  1983.3& $2021.41\pm0.05$ &  $2007.67 \pm 0.02 $    \\
$a$ (AU)         &  6.3       &  -               & $4.93 \pm 0.01 $   \\
$e$              &   0       & $0.2436\pm0.0012$ &     $ 0.245 \pm 0.001 $  \\
$\omega_0$ (deg) &   0       & $158.81\pm0.62$ &     $ 157.5 \pm 0.6 $   \\
$\Omega_1$ (deg) &   16      & $56.13\pm0.17$ &    $ 56.25 \pm 0.17 $   \\
$i$ (deg)        &   144     & $130.79\pm0.20$ &   $ 130.3 \pm 0.3 $   \\
$M_r$ ($\frac{M_2}{M_{tot}}$)& - & -           &  $ 0.140 \pm 0.008 $  \\
$V_0$ (km/s)     &   -         &  -             &   $ -12.307 \pm 0.04 $    \\
$V_1$ (km/s)     &   -         &  -             &   $ -0.07 \pm 0.11 $    \\
\hline
$M_{tot} $ ($M_{\astrosun}$) &   - & $0.666\pm0.035$ &   $ 0.642 \pm 0.005 $    \\
$M_{2} $ ($M_{\astrosun}$) &   - & - &   $ 0.090 \pm 0.005 $    \\
$M_{1} $ ($M_{\astrosun}$) &   - & - &   $ 0.553 \pm 0.007 $    \\

\hline
\end{tabular}
\end{table*}

\begin{figure}
    \begin{center}
      \includegraphics[width=0.5\textwidth]{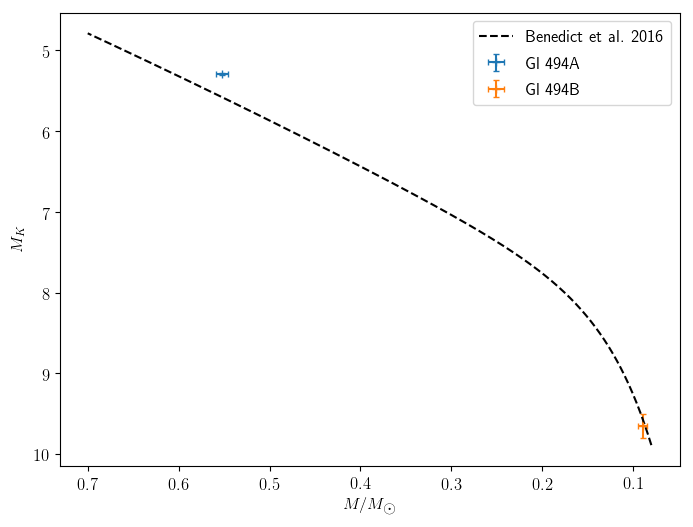}
      \caption{Comparison of the mass and absolute magnitudes obtained from this work with the mass-luminosity relationship for the low mass stars in the solar neighborhood, showing reasonable agreement}.
      \label{fig:benedict-plot}
    \end{center}
  \end{figure}

  The apparent magnitude of the system from 2MASS $m_K = 5.578 \pm0.016$ together with the distance and contrast provide us with absolute magnitude of $M_{K} = 5.29 \pm0.017$ for Gl 494A and $M_{K} = 9.65 \pm0.15$ for Gl 494B. Figure \ref{fig:benedict-plot} shows how this compares to the expected mass-luminosity relationship provided by \citet{Benedict2016} for the M dwarfs in the solar neighborhood. The two components are found to be in reasonable agreement with the model.

\section{Conclusion}
 
  We have demonstrated using both simulation and on-sky data that it is possible to use our knowledge of the PSF to recover the saturated pixels of images. Unlike the original images, the mended images can then be used successfully in kernel-phase analysis. The proposed method works particularly well in the case of images dominated by the PSF of a single star, like in the case of high contrast binary stars or exoplanetary systems. Its primary purpose is to expand the field of applicability of kernel-phase methods, especially to the larger body of archival space telescope images.\par
  As an example, we have used the technique on images from the HST of the Gl 494AB system with a saturated core, and produced a new visual astrometry data point at an interesting older epoch and small uncertainties. Combining this information with radial velocity data, we have improved the orbital model of the system and determined the masses of its two components.\par
  As remarked by \citet{Torres1999}, providing high precision visual measurement of astrometric and spectroscopic binary systems is key to obtain accurate orbital parameters and masses, and this work shows once more how kernel analysis can play a major role in the follow-up of the binary systems that will be discovered in large quantities by the Gaia mission. The possibility of using saturated images could lead us to reevaluate how to optimize the observation strategy for this goal, as well as for the discovery of new stellar binary systems, especially in the context of the James Webb Space Telescope. 

\begin{acknowledgements}
The authors wish to thank Xavier Delfosse, Stephane Udry, Damien Ségransan and the CORAVEL team at the observatoire astronomique de l'université de Genève for allowing us to use the data from the CORAVEL instrument; and Sébastien Peretti for helping us with the data. \par

KERNEL has received funding from the European Research Council (ERC) under the European Union's Horizon 2020 research and innovation program (grant agreement CoG - 683029).

This publication makes use of data products from the Two Micron All Sky Survey, which is a joint project of the University of Massachusetts and the Infrared Processing and Analysis Center/California Institute of Technology, funded by the National Aeronautics and Space Administration and the National Science Foundation.
\end{acknowledgements}

\bibliographystyle{aa}
\bibliography{SatRecovery}

\end{document}